\documentclass[12pt,a4paper]{article}

\usepackage[english]{babel}
\usepackage{psfrag,graphicx}
\usepackage{amsmath}

\newcommand{\G}{\widetilde{G}}
\newcommand{\intp}{\frac{d^4 p}{(2\pi)^4}}
\newcommand{\dk}{\frac{d^4 k}{(2\pi)^4}}
\newcommand{\dq}{\frac{d^4 q}{(2\pi)^4}}

\newcommand{\de}{\delta}

\newcommand\pubnumber{UAB-FT-553}%\\PSI-PR-??-??\\UB-ECM-PF-03/13}

\newcommand\pubblock{\rightline{\begin{tabular}{l} \pubnumber \end{tabular}}}

\title{Radiative corrections in 5D and 6D \\ expanding in winding modes}
\author{Leandro Da Rold\footnote{daroldl@ifae.es}\\
\small{IFAE, Universitat Aut{\`o}noma de Barcelona, 08193 Bellaterra
  (Barcelona), Spain}}

\begin{document}
\maketitle
\begin{abstract}
We compute radiative corrections in five and six dimensional field
theories, using winding modes in mixed momentum-coordinate space. This
method provides a simple way of finding UV divergencies, finite corrections and
localized terms when the space is compactified on orbifolds. As an
application we compute the finite piece of scalar masses, the
logarithmic contributions
to the couplings and the effect of localized parallel and
perpendicular kinetic terms. We apply it
to get a two loop effective potential that can stabilize large extra
dimensions.
\end{abstract}
\pubblock
\newpage
\section{Introduction}
The Standard Model is not a fundamental theory and
there have been many proposals to go beyond it. It is expected that a
field theory with extra dimensions arises as the low energy limit of a
fundamental string theory. For this reason extra dimensions
are a common feature of any theory valid at high energies.
Almost all problems of particle physics have been
reformulated in this context giving new possibilities. In
particular, physics of large extra dimensions
can provide solutions to the hierarchy problem \cite{add}, \cite{rs} and new
mechanism of symmetry breaking \cite{simbreak}, that can
be tested in the next high energy experiments. 

In this note we developed a formalism that allows us to compute loop
corrections in field theories with large extra dimensions, separating UV
divergencies from finite contributions, in a direct product
space. Usually, higher dimensional theories are formulated using
Kaluza-Klein decomposition. Instead in this work we will use winding
modes. In a five dimensional theory these modes are obtained
propagating around the circle of the extra dimension. Two paths with
different windings are topologically different. Ultraviolet
divergencies in loop integrals are associated to the zero winding
mode. Non zero modes are long distance and each of them will give
finite terms~\cite{radEW}. Then this formalism is very useful to
separate finite from divergent contributions.

We will use winding decomposition to compute radiative
corrections on 5D and 6D theories. When the theory is compactified on an
orbifold translation symmetry is broken on the fixed points; we will show
how localized terms are generated at one loop. 
We also analyze the role of localized kinetic
terms, parallel and perpendicular to the brane, interpreting them in a
very intuitive way, and showing their physical effects. We show these
terms can provide new mechanisms of symmetry breaking. We will compute
general finite masses and also self and gauge couplings using
this method. 

When there are more than one extra dimensions, winding modes prove to
be very useful. In particular, we consider a 6D theory and after
obtaining the propagator in mixed representation,
we use it to compute radiative corrections, the finite piece of the
masses and the couplings.

As an interesting application we will apply this formalism to compute
a two loop effective potential. We will prove that this potential can
stabilize large extra dimensions when there are brane terms.

In section 2 we define winding
modes working on a mixed momentum-coordinate representation. We apply
this idea to a 5D theory in section 3. In section 4 we show
how to work with more than one extra dimension. In section 5
we compute a two loop effective potential and section 6 is for conclusions.

During the writing of this work another paper appeared~\cite{suizos} on similar
subjects, reaching the same results about mixed momentum-coordinate
representation and mass terms.
\section{Winding modes}\label{windings}
Let's consider a 5D theory compactified on $\mathcal{M}^4\times
\mathcal{C}^1$, where $\mathcal{M}^4$ is 4D Minkowsky space and
$\mathcal{C}^1$ is a compact 1-D manifold. 
If we can write $\mathcal{C}^1=\mathcal{R}^1/G$, where
$\mathcal{R}^1$ is the real line and $G$ is a discrete group
acting freely on $\mathcal{R}^1$, then we can define winding
modes. The simplest example is $\mathcal{C}^1=S^1$ (a circle) and $G=\mathcal{Z}$ (the set of integer numbers, with
the sum defined as the group product). In this case we obtain the
compact space identifying $y\sim y+n 2\pi R$. Due to the
identification $y\in[0,2\pi R)$, the index $n$ labels winding
modes. 

This idea suggests the following procedure: we compute on the
infinite space and identify $y\sim y+n 2\pi R$ to
obtain the physical magnitudes on the compact space. For a general
compact space $\mathcal{C}^m=M^m/G$ of dimension $m$ ($M^m$ a non compact
m-dimensional manifold), we identify
$z\sim g(z)$, $z\in \mathcal{C}^m$ and $g\in G$. Then we can define
winding modes for every $g$, and associate divergencies to the zero
mode $g=1$.

To show how this formalism works we will compute the massless scalar propagator on
Euclidean 5D spacetime. We work on a mixed representation
$(p_{\mu},y)$, where $p_{\mu}$ is the 4D momentum and $y$ the
coordinate in the extra dimension. Then we can obtain the Green's
function from the following equation
\begin{equation}\label{mixta}
(p^2-\partial^2_y)\G(p;y-y')=\delta(y-y'),
\end{equation}
\noindent where we have Fourier transform to momentum space only
on the 4D space, and $p^2=p^{\mu}p_{\mu}$. Solving this
equation we get
\begin{equation}\label{GR1}
\G(p;y-y')=\frac{e^{-p|y-y'|}}{2 p}.
\end{equation}
\noindent Due to translation invariance we see that the propagator only
depends on $|y-y'|$. If we consider a massive scalar field we just
have to replace $p^2\rightarrow p^2+m^2$.

\begin{figure}[htbp]
  \centering
  \psfrag{0}{\small{$0$}}
  \psfrag{1}{\small{$1$}}
  \psfrag{n}{\small{$n$}}
  \psfrag{y}{\small{$y$}}
  \psfrag{y'+w}{\small{$y'\!+2\pi R$}}
  \psfrag{y'}{\small{$y'$}}
  \includegraphics[height=1.5cm, width=8cm]{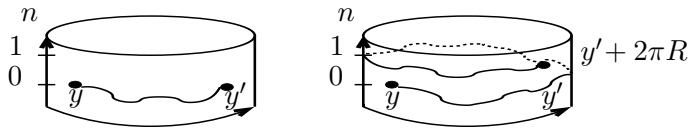}
  \caption{\textit{Two equivalent contributions for the propagator between
  $y$ and $y'$. In the vertical axis we represent the number of
  windings.}}
  \label{Figwindings}
\end{figure}

To get the propagator on the compact space we identify $|y-y'|\sim
|y-y'+2n\pi R|$ (see fig.~\ref{Figwindings}). Then we
restrict $y,y'\in [0,2\pi R)$ and sum over
windings
\begin{equation}
\G^{cir}(p;y,y')=\sum_{n=-\infty}^{n=\infty}\G(p;y-y'+2 n \pi R)=\sum_n
\frac{e^{-p|y-y'+ 2 n \pi R|}}{2 p} .
\end{equation}
\noindent We can solve this sum, but it is more useful
to consider the contribution of each mode. In the last equation we
can see that for $n\neq 0$ the propagator is exponentially damped,
therefore loop integrals over momenta will be finite. On the other hand, for
$n=0$, the propagator goes as $p^{-1}$ when evaluated in $y=y'$.
For these reasons, compactifying on a circle, we will obtain
divergent contributions for the winding 0-modes, and finite
contributions for the other modes.
Therefore, using the winding modes
formalism, is very easy to separate divergent from finite
contributions. The divergent terms are associated to short
distances, that is why the divergencies resulting from the
0-mode are the same as the divergencies of the uncompactified
theory. For the masses, the radiative corrections are dominated by
high energy effects. High energy implies small distances, then for the
masses, the small $n$ contributions dominate over large $n$. This is
not the case when we do Kaluza-Klein decomposition. This is one of the
advantages of winding decomposition.

\bf{Orbifold compactification}\\
\rm
Orbifolds are used to obtain 4D chiral fermions from a higher
dimensional theory. In general we can obtain an orbifold with a
discrete group $F$ acting non freely on the compact space
$\mathcal{C}$. Points of $\mathcal{C}$ left invariant by $F$ are
fixed points, and there $\mathcal{C}/F$ is singular. The simplest
example is the orbifold $S^1/Z_2$, where $Z_2: y\rightarrow -y$ is
the parity transformation on the extra dimension. Due to the
identification $y\in[0,\pi R]$ and the fixed points are $y=0,\pi
R$. To give a complete description we also have to specify field parities.

We first consider a scalar field on $S^1/Z_2$, with definite parity $Z_2
\phi(x^{\mu},y)=\phi(x^{\mu},-y)=\pm \phi(x^{\mu},y)$. Due to the
identification $y\sim -y$, we can propagate from $y$ to $y'$ and from
$y$ to $-y'$, then the propagator is~\cite{Alex1}
\begin{equation}\label{GZ}
\G^{orb}_{\pm}(p;y,y')=\sum_n (\frac{e^{-p|y-y'+2 n \pi R|}}{2
    p}\pm \frac{e^{-p|y+y'+ 2 n \pi R|}}{2 p}),
\end{equation}
\noindent where $y,y'\in[0,\pi R]$, and $\pm$ depends on the field
parity. This propagator depends on $(y+y')$ due to the breaking of
translation invariance.

In this equation we can see that the propagator goes as $p^{-1}$
for the following limits of the coordinates and windings:
$(y\rightarrow 0,y'\rightarrow 0,n=0)$ and $(y\rightarrow \pi
R,y'\rightarrow \pi R,n=-1)$. Then we expect divergencies
localized on the fixed points of the orbifold~\cite{hailu}. We can expect this
terms because they don't break any symmetry of the theory.

\section{5D radiative corrections on $\lambda \phi^4$}\label{5Drad}
\begin{figure}[htbp]
    \centering %\setlength\unitlength{.8cm}
    \psfrag{l}{$\lambda$}
    \includegraphics[height=2cm]{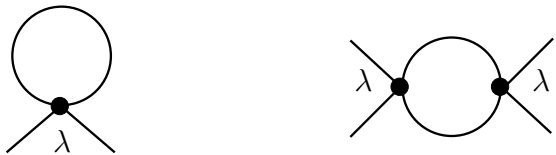}
    \caption{\textit{Feynman diagrams for one-loop mass and vertex in the scalar interacting theory.}}
    \label{Fig:lazos5D}
\end{figure}

We compute one-loop radiative corrections of the scalar interacting theory
defined by 
\begin{equation}\label{teorialambdaphi4}
S=\int d^4x\,
dy\,[\frac{1}{2}(\partial_M\phi)^2-\frac{\lambda}{4!}\phi^4].
\end{equation}
\noindent We will consider our toy model an effective theory valid to some
cut-off $\Lambda$, that by naive dimensional analysis should be
$\Lambda \sim 24 \pi^3 \lambda^{-1}$. Furthermore, the lagrangian in
eq.~(\ref{teorialambdaphi4}) shall be defined as the effective theory
at the scale $\Lambda$. In performing quantum
corrections we will cut off the 4D momentum integral at the scale
$\Lambda$. The loop contributions that depend on $\Lambda$ will signal
the divergencies of the 5D theory.
In fig.~\ref{Fig:lazos5D} we can see the Feynman diagrams
that renormalize the two point function and the coupling at one loop.
Then we can write the effective action with one-loop quantum
corrections as
\begin{equation}\label{Seff}
S_{eff}=S_{cl}+S_2+S_4+...=S_{cl}-\int dy \;m^2(y)\phi^2(y)-\int dy dy' \phi^2(y)\lambda(y,y')\phi^2(y')+...
\end{equation}
\noindent Let's consider first, the extra dimension
compactified on a circle, then $y\in(0,2\pi R)$. For the circle
$m^2(y)$ is constant and is given by (the one-loop mass
was calculated in Ref.~\cite{DPQ} using K-K decomposition)
\begin{equation}\label{0masaS1}
m^2_{cir}=\frac{\lambda}{2}\int \intp
(\frac{1}{2p}+\sum_{n\neq0} \frac{e^{-p2|n|\pi R}}{2p})=
\frac{\lambda}{16 \pi^2} (\frac{\Lambda^3}{6}+\frac{1}{(2\pi
R)^3}\sum_{n \neq 0} \frac{1}{|n|^3}),
\end{equation}
%\begin{equation}\label{0masaS1}
%S_2=-\frac{\lambda\phi^2_c}{2}\int dy \int \intp
%(\frac{1}{2p}+\sum_{n\neq0} \frac{e^{-p2|n|\pi R}}{2p})=
%-\frac{\lambda\phi^2_c}{16 \pi^2} (\frac{\Lambda^3 \pi R}{3}+\frac{1}{(2\pi
%R)^2}\sum_{n \neq 0} \frac{1}{|n|^3}),
%\end{equation}
\noindent where we have thrown away terms that cancel when
$\Lambda R\rightarrow \infty$. As we argued in the previous section,
divergencies are associated to $n=0$ and finite terms to $n\neq
0$. This expresion is valid for general $\phi(y)$, because $m^2(y)$ is
independent of $y$.

If there is a symmetry (like supersymmetry or a gauge symmetry) prohibiting
divergent masses, then the finite term is a prediction of the theory.

Now we compute $S_4$ at one loop. For simplicity, we will take
$\phi(y)=\phi_c=$constant. According to this, expanding in powers of external momenta we
just keep the zero order terms. Then we get for $S_4$
\begin{equation}\label{Loop}
S_4=-\frac{\lambda^2\phi^4_c}{2} \int dy\int dy'\sum_{n,n'}\int \intp \G(p,y-y'+2n\pi R)\G(p,y'-y+2n'\pi R).
\end{equation}
\noindent %where $\delta \lambda^{4D}_{S^1}$ is the 4D one-loop coupling because we have integrated over the extra dimensions. 
There are two propagators involved, so there are two
winding indexes. When the topology of the extra space is more
complicated (for example when there are more than one extra
dimension) it is useful to express this equation in terms of just
one propagator as 
\begin{equation}\label{I}
S_4=-\mathcal{I}\lambda^2\phi_c^4/2,
\end{equation}
\noindent where $\mathcal{I}$ is given by~\footnote{To
  see that eq.~(\ref{Loop}) and eq.~(\ref{I}) are the same we can write
  eq.~(\ref{Loop}) in K-K modes without external momenta as
  $\sum_{p_5}\int \intp \frac{1}{(p^2+p_5^2)^2}$, with $p_5$ the
  momentum in the extra dimension. The integrand can be written as
$\frac{1}{(2\pi)^4}\frac{-d}{dp^2}(p^2+p_5^2)^{-1}$, and by Fourier
transformation we obtain eq. (\ref{I}).}
\begin{equation}\label{simple2}
\mathcal{I}=-\int dy  \sum_{n}\int \intp
\frac{d}{dp^2}\G(p;y,y+2n\pi R).
\end{equation} 
Integrating coordinates and momenta we get the desired result, and the
one-loop contribution to $\lambda(y,y')$ is 
\begin{equation}\label{lambdaS}
\lambda_{cir}=\frac{\lambda^2}{64\pi^2}
(\Lambda-\sum_{n\neq 0}\frac{1}{|n|\pi R}),\;\;\;\;\;\;\;\;\;\;\phi=\phi_c.
\end{equation} 
\noindent The sum over $n$ is logarithmically divergent in the IR,
so we have to introduce an IR cut-off, that means that we sum to
$n_{max}=(2\pi R \mu_{ir})^{-1}$, regulating the long distance behavior. If the
field is massive the mass is the natural cut-off. 
The IR logarithm is the same as in 4D, and this can be easily understood in
terms of K-K modes. The zero K-K mode is massless, and this is the mode that
propagates long distances. 

\subsection{Radiative corrections on orbifolds}\label{orbren}
The one-loop contribution to $S_2$ on $S^1/Z_2$ are generated by the
following expression (in Ref.~\cite{schmaltz} are given the masses for
K-K modes)
\begin{equation}\label{masaZ1}
S_2=-\frac{\lambda}{2} \int_0^{\pi R} \!\!dy \; \phi^2(y) \int \intp
\sum_n (\frac{e^{-p2|n|\pi R}}{2p}\pm\frac{e^{-p2|y+n\pi
    R|}}{2p})\equiv S_S \pm S_Z,
\end{equation}
\noindent where $S_S$ is the same as $S_2$ for the circle, except that
$y\in(0,\pi R)$. The second term depends on $y$ and it's new. As we argued, it
has divergencies for $n=0,-1$. 
%We consider $\phi$ constant and $S_Z$ is given by
%\begin{equation}\label{masaZ2}
%S_Z=- \frac{\lambda\phi_c^2}{16\pi^2} \big[\sum_{n\neq 0,-1}
%\frac{(1+2n)sgn(n)}{n^2(1+n)^2 16\pi R}+\frac{\Lambda^2}{4}-\frac{1}{8
%  \pi R}\big],
%\end{equation}

To obtain the divergencies we expand
$\phi^2(y)$ in powers of $y$ around the fixed points
$y_{fp}$. Expanding $\phi^2$ to second order, $S_Z$ is given by
\begin{equation}\label{S2Z}
S_Z\simeq -\frac{\lambda}{16\pi^2}
\sum_{fp}\big\{\phi^2(y_{fp})\big[\sum_{n\neq 0,-1}
\frac{(1+2n)sgn(n)}{n^2(1+n)^2 32\pi^2 R^2}+
\frac{\Lambda^2}{8}-\frac{1}{16 \pi^2 R^2}\big]
+\partial_y^2\phi^2(y_{fp}) \frac{\log(\Lambda R)}{16}\big\}.
\end{equation}
\noindent The $\Lambda^2$ divergencies are associated to $n=0$ when
$y\rightarrow0$ and to $n=-1$ when $y\rightarrow\pi R$. Then these
divergencies are localized on the fixed points of the orbifold. This
is because the orbifold compactification breaks translation invariance
in these points. 

From eq.~(\ref{masaZ1}) we can see that $m^2(y)$ has two
contributions. The first one is the same as in the circle,
eq.~(\ref{0masaS1}). The second contribution depends on $y$ and, from
eq.~(\ref{S2Z}), we can see the divergent terms. We split $m^2(y)$
in divergent and finite contributions as
$m^2(y)=m^2_{div}(y)+m^2_f(y)$. Then the divergent contribution
is given by
\begin{equation}
m^2_{div\pm}(y)=\frac{\lambda}{16\pi^2}
\big\{\frac{\Lambda^3}{6}\pm\sum_{fp}\de_{fp}[\frac{\Lambda^2}{8}+\frac{\log(\Lambda R)}{16}\partial_y^2]\big\}.
\end{equation}
\noindent The finite term can be computed for a constant field, and
is given by
\begin{equation}
m^2_{f\pm}=\frac{\lambda}{128\pi^5 R^3}\big[\sum_{n\neq 0}\frac{1}{|n|^3}\pm\sum_{n\neq 0,-1}
\frac{(1+2n)sgn(n)}{2n^2(1+n)^2}\mp 1\big],\;\;\;\;\;\;\;\;\;\;\phi=\phi_c.
\end{equation}

To get $S_4$, we again expand the fields in powers of $y$ around the
fixed points $y_{fp}$. We just take the zero order term in the power
series and use eqs.~(\ref{I}) and~(\ref{simple2}) with the orbifold
propagator. Then $S_4$ is given by
\begin{equation}
S_4\simeq-
\frac{\lambda^2}{16\pi^2}\sum_{fp}\phi^4(y_{fp})\big[\frac{\Lambda \pi R}{8}+\sum_{n>1}(\frac{1}{4n}\pm\frac{1}{8}
\log\frac{n+1}{n-1})\pm \frac{\log(\Lambda R)}{4}\big].
\end{equation} 

\noindent The linear UV divergence is due to the zero winding
mode. The logarithmic divergence is localized on the fixed points,
then it is 4D, and is
associated to $n=0,-1$. Therefore we can write $S_4$ as
\begin{equation}\label{S4Z}
S_4=-\int dy\;\lambda^{\pm}(y)\;\phi^4(y),
\end{equation}
\begin{equation}\label{LoopSZ} 
\lambda^{\pm}(y)=\lambda_f^{\pm}(y)+\frac{\lambda^2}{64\pi^2}
\sum_{fp}\de_{fp}[\frac{\Lambda\pi R}{2}\pm\log(\Lambda R)],
\end{equation}
\noindent where $\lambda_f^{\pm}(y)$ is a finite coupling. For a constant
field $\lambda_f^{\pm}$ is given by 
\begin{equation}
\lambda_f^{\pm}(y)=\frac{\lambda^2}{16\pi^2}\big[\frac{1}{\pi
  R}\sum_{n>1}(\frac{1}{2n}\pm\frac{1}{4}
\log\frac{n+1}{n-1})\big],\;\;\;\;\;\;\;\;\;\;\phi=\phi_c.
\end{equation}
\noindent If the field is even there are
logarithmic IR divergences, as in the circle, but this doesn't
happen in the odd case. This is easier to understand in terms of K-K
modes: only the even field has a massless mode.

In dimensional regularization scheme, the effective scale $\Lambda$ in
the 4D logarithm of the coupling, eq.~(\ref{LoopSZ}), is replaced by an
arbitrary scale $\mu$. Further discussions in this scheme can be found
in Ref.~\cite{hailu}.

\subsection{Localized kinetic terms}
We have seen that new terms localized on the branes have been induced,
showing that these terms should be taken from the begining. Here we want to
consider the effect of these terms on the physical parameters. We
are interested in the kinetic terms: they can be parallel
$\mathcal{L}^{\parallel}=a p^2 \de_{fp}\phi^2$ or perpendicular
$\mathcal{L}^{\perp}=b \de_{fp}\partial_y^2\phi^2$ to the brane (see
\cite{perezvictoria} for the most general case), where $a,b$
are the couplings. Perpendicular kinetic terms generate classical
divergencies, similar to classical divergencies in
electromagnetism. They can be regularized with a fat brane, for
example 
defining a $\delta_{\epsilon}\rightarrow \delta$ when $\epsilon
\rightarrow 0$. Then, including the right counterterms, we can
renormalize the theory (for classical renormalization with branes of
codimension bigger than one see~\cite{braneflows}).

\begin{figure}
  \centering
  \psfrag{G}{$\G^{\perp}=$}
  \psfrag{+}{$+$}
  \psfrag{...}{$...$}
  \psfrag{b}{$b$}
  \includegraphics[width=8cm]{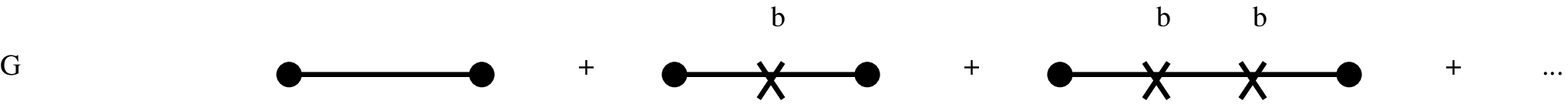}
  \caption{\textit{Perturbative expansion of the scalar propagator with
  perpendicular kinetic brane terms.}}
  \label{figperturbaciones}
\end{figure}

The renormalization can be performed in the following way: we obtain
the propagator
$\G^{\perp}$ (with perpendicular terms) as a perturbative series with
$b$-vertex insertions, as is shown in
fig.~\ref{figperturbaciones}. Every term of order $n$, has
divergent contributions of order $\epsilon^{-n}\sim\Lambda_B^n$, where
$\Lambda_B$ is a brane scale.
%(we have done $\epsilon\sim\Lambda^{-1}$ because we consider the
%theory above our model solves the thin brane). 
To cancel the divergent terms when $\epsilon \rightarrow 0$, we have
to add counterterms proportional to $\de^n$, where $n$ is the number
of vertex insertions. The divergencies can be interpreted as
contributions coming from processes at energy $\Lambda_B$ on the
branes. Then, including these counterterms, we are neglecting high
energy contributions that can feel the brane structure. But this is
exactly what we want, an approximation valid for energies $E\ll
\Lambda_B$. 
%All these divergent terms would produce a jump for the propagator
%evaluated on the brane, allowing for field discontinuities
%there. Then our prescription is to consider continuos fields, as far
%as there are only low energies $E\ll \Lambda_B$. Of course this is
%the regime where all our theory has sense.

After that, in the theory there are only parallel terms. To see this we
can resum the perturbative series (without the divergent terms)~\cite{perezvictoria} and get
\begin{equation}
\G^{\perp}(p;y,y')=\G^{(0)}(p;y,y')-\frac{b p^2
  \G^{(0)}(p;y,y_j)\G^{(0)}(p;y_j,y')}{1+b p^2\G^{(0)}(p;y_j,y_j)}.
\end{equation}
\noindent where $\G^{(0)}$ is the free propagator. Then $\G^{\perp}$ is the same propagator as $\G^{\parallel}$ (with
parallel kinetic terms) changing $b\rightarrow a$, as we will see
in the following paragraphs, eq.~(\ref{Gkinetic}). So there is nothing new considering
perpendicular couplings, we can obtain all the relevant information
analyzing parallel terms.  For these reasons we will concentrate on
parallel kinetic terms.

\bf{Parallel kinetic terms}\\
\rm
Let's consider a scalar interacting theory
\begin{equation}
S=\int d^4x dy
[\frac{1}{2}(\partial_M\phi)^2+\sum_{fp}\frac{a_{fp}}{2}\de(y-y_{fp})(\partial_{\mu}\phi)^2-\frac{\lambda}{4!} \phi^4].
\end{equation}
To obtain the propagator we apply the ideas of the previous
sections. For the moment we suppose that the extra dimension is
infinite and compute the propagator with just one delta on one of the
fixed points $y=y_{fp}$. In this case the Green's function
equation in mixed momentum-coordinate representation is
\begin{equation}
(p^2-\partial^2_y)\G(p;y,y')+a p^2\delta(y-y_{fp})\G(p;y_{fp},y')=\delta(y-y').
\end{equation}
This equation is similar to the free one but with a new source of
magnitude $-ap^2 \G(p;y_{fp},y')$ in $y=y_{fp}$. Then the propagator is
\begin{equation}\label{Gkinetic}
\G(p;y,y')=\frac{e^{-p|y-y'|}}{2 p}-\frac{ap}{2+ap}\frac{e^{-p(|y-y_{fp}|+|y_{fp}-y'|)}}{2 p}.
\end{equation}
It is immediate to read the second term as a reflection of
magnitude $\mathcal{X}=ap/(2+ap)$ on the brane on $y_{fp}$. Once we
realize this, we can compute the propagator in the compact space in
a perturbative way (in \cite{radEW} we can see the series with mass insertions
ressumed, but we want to keep our intuition working with winding
modes). We have to sum over all the contributions coming from
reflections on $\de(y-y_{fp}+2n\pi R)$, in a similar way we do
with light travelling in a medium with different indexes of
reflection $\mathcal{X}$ and transmission
$\mathcal{T}=1-\mathcal{X}$. A wave of amplitude $\mathcal{A}$
arrives to a $\de$ in $n\pi R$, there
$\mathcal{A}\mathcal{X}$ is reflected and $\mathcal{A}\mathcal{T}$ is transmited,
and due to the propagation the wave amplitude is damped $e^{-pd}$
after travelling a distance $d$. With these rules we can obtain the
propagator to any order in $e^{-p\pi R}$.

We will study the limit of $a\rightarrow \infty$ for fixed 4D coupling
$\lambda^R_{4D}=\lambda \pi R/(\pi R+a)^2$. The propagator in the
orbifold (with 4D kinetic term canonically normalized) is given by
\begin{equation}\label{Gkinetic3}
\G^R_{\pm}(p;y,y')= \sqrt{\lambda^R_{4D}}\frac{\pi R+a}{\pi R}[\G(p;y,y')\pm\G(p;y,-y')],
\end{equation}
\noindent where $\G(p;y,y')$ is given in eq.~(\ref{Gkinetic}).

For simplicity we consider first a kinetic term on $y_{fp}=0$. The
propagator of eq.~(\ref{Gkinetic}) has two regimes depending on
wether $ap\gg 1$ or $ap\ll 1$. We define a critic winding
$n_c=a(2\pi R)^{-1}$, then $|n|\ll n_c$ corresponds to the high energy
regime and $|n|\gg n_c$ to the low energy. Therefore the propagator is given by
\begin{equation}\label{Gkinetic2}
\begin{aligned}
&\G^{n<n_c}_{\pm}(p;y,y')\simeq\sqrt{\lambda^R_{4D}}\frac{\pi R+a}{\pi
  R} \G^{orb}_{-}(p;y,y'),%\;\;\;\;\;\;\;\;\;
\\ &\G^{n>n_c}_{\pm}(p;y,y')\simeq \sqrt{\lambda^R_{4D}}\frac{\pi R+a}{\pi R}\G^{orb}_{\pm}(p;y,y').
\end{aligned}
\end{equation}
%\sqrt{\frac{\lambda}{\pi R}} 
Thus we see that at short distances the field is similar to an odd
field. When $n\sim n_c$ $\G$ is more complicated, but for a
rough estimation we can consider eq.~(\ref{Gkinetic2}) valid for
all $n$. It is important to note that the two pieces of equation
(\ref{Gkinetic2}) are sensible to different values of momenta. 

\bf{Effect of parallel terms on the physical parameters}\\
\rm
Once we have the propagator we can compute the mass and
selfcoupling at one-loop with brane kinetic terms.
Repeating the steps done in the previous section we calculate
the one-loop mass. We consider a constant field $\phi_c$, and integrating over
the extra dimension we obtain the 4D mass. There are divergent and
finite contributions. Here we show the finite terms, splitted in two
contributions, depending on the winding values
\begin{equation}\label{Map}
\begin{aligned}
m^2_{4D\pm}&=\frac{\lambda^R_{4D}} {128\pi^4 R^2}\Big[\sum_{n=1}^{n_c-1}e^{-n/nc}\frac{n^2+2n n_c+2n_c^2}{n^3n_c^2}-e^{-1/nc}\frac{1+n_c}{n_c}+\frac{2e^{-1}}{n_c^2}\\
%\frac{\lambda^R_{4D}}{16\pi^2}[\frac{\Lambda^2}{2}+\frac{1}{2\pi^2R^2}\sum_{n=1}^{n_c}\frac{1}{n^2}\pm2(\frac{\Lambda}{2\pi
%  R}-\frac{1}{4\pi^2 R^2}+\frac{1}{4\pi^2 R^2}\sum_{n=1}^{n_c}\frac{1}{n+n^2})],\\
&+ %\frac{\lambda^R_{4D}\frac{\pi R+a}{\pi R}}{128\pi^4 R^2} 
\sum_{n=n_c}^{\infty}(-e^{-n/nc}\frac{n^2+2n
    n_c+2n_c^2}{n^3n_c^2}+\frac{1}{n^3}\pm
    \frac{1+2n}{n^2(n+1)^2})+\frac{2e^{-1}}{n_c^2}(1\mp 1)\Big],
%\frac{\lambda^R_{4D}n_c}{16\pi^2}[\frac{1}{4\pi^2R^2}\sum_{n>n_c}^{\infty}\frac{1}{n^3}\pm\frac{1}{8\pi^2R^2}\sum_{n>n_c}^{\infty}\frac{1+2n}{n^2(1+n)^2}].
\end{aligned}
\end{equation}
\noindent where the subindex $\pm$ is for the even and odd
contribution. Then eq.~(\ref{Map}) gives the one-loop finite
contribution to the 4D mass, with brane kinetic terms. 

We can compare the finite 4D mass term with brane kinetic couplings with
the one without them. To do this we consider that the 4D couplings of
both theories are the same (for constant field $\phi_c$). Then we get
\begin{equation}\label{m4D}
\begin{aligned}
&m^2_{4D\pm}\rightarrow\frac{\lambda_{4D}}{16\pi^4
  R^2}\frac{\zeta(3)\pm1/2}{8},\;\;\;\;\;\;\;\;\;\;\;\;\;\;\;\;\;\;\;\;\;\;\;\;\;\;\;\;\;\;\;\;\;\;\;\;\;\;\;\;\;\;\;\;\;\;\;\;\;\;\;\;\;\;\;a\rightarrow 0,
\\ &m^2_{4D\pm}\simeq \frac{\lambda_{4D}}{16\pi^4
  R^2}[\frac{\zeta(3)-1/2}{8}+\frac{\alpha_{\pm}}{n_c^2}+O(1/n_c^3)],\;\;\;\;
\;\;\;\;\;\;\;\;\;\;\;\;\;\;\;\;\;\;\;\;\;\;\;\;a\rightarrow \infty,
\end{aligned}
\end{equation}
\noindent where $n_c=a/(2\pi R)$ and
\begin{equation}
\alpha_{\pm}=3/2-4/e,0.
\end{equation}
\noindent In eq.~(\ref{m4D}) we have put the first correction in powers of
$1/n_c$. The odd field doesn't coupled to the brane at $y=0$, that's
why the mass doesn't change for this mode.

Now we compute the vertex with constant field $\phi_c$. First we
discuss the high energy regime. As we said before, in this regime the
field seems an odd field. Then we get the linear and logarithmic
dependence on $\Lambda$ in the same way we did in the previous section.

Second, we consider the low energy regime that corresponds to
windings $n>n_c$ in both propagators involved in the Feynman
diagram. This case is the same as the one without brane terms, but
summing over windings $n>n_c$. If the field is even under $Z_2$,
there is an IR logarithm. Then, the finite and logarithmic one loop
contribution to the 4D coupling are given by
\begin{equation}
\lambda^{+}_{4D}\simeq \frac{\lambda^2_{4D}}{16\pi^2} [C-\log(n_c
R\mu_{ir})-\frac{1}{2}\log(\Lambda R)],
\end{equation}
\noindent where $C\sim-1$ and the first logarithm is the IR long
distance. For an odd field we get a similar result without the IR logarithm.

Let's discuss now what happen with brane kinetic terms in
$y_{fp}=\pi R$. In this case the limit of $a\gg R$ corresponds to
an opaque brane, and the reflection has a minus sign. Then the effect
of this brane is again the same as considering an odd field. So we can
get an odd field puting (almost) opaque branes. This suggests a new way o
symmetry breaking: let's suppose that the components of a multiplet
have different brane couplings. Then the effect of
these couplings will be the same as choosing different boundary conditions for
the fields of a given multiplet, breaking the symmetry under which the
multiplet transforms.

\subsection{One-loop gauge coupling}
We consider as an application of the previous formalism, a 5D theory
with gauge fields and a scalar charged field, transforming with
representation S. 
We want to get the logarithmic divergencies of the gauge coupling, due
to the scalar fields, in an orbifold. Then we
consider the one loop scalar contribution to the vacuum polarization
$\Pi_{MN}$. This have been computed with K-K modes
\cite{5to4}, here we get the same result with winding modes.

We consider the effective 4D theory, that is the theory obtained after
integration over the extra dimension with constant
fields (zero K-K modes). We define $g_0$ as the effective 4D coupling
of an abelian theory, at the scale $\Lambda$.
Then, after some manipulations, we can write the divergent part of the
one-loop vacuum polarization as
\begin{equation}\label{pi}
\Pi(k^2=0)=\frac{g_0^2}{3}\sum_{q_5} \int \dq
\frac{1}{(q_4^2+q_5^2)^2}=-\frac{g_0^2}{3}\int dy\int \intp
\frac{d}{dp^2}\G^{orb}_{\pm}(p;y,y).
\end{equation}
\noindent The momentum
integral is a loop with two scalar propagators without external
momenta, then it is the same as the one loop $\lambda$ contribution.
On the r.h.s. of eq.~(\ref{pi}) we have written the vacuum
polarization in terms of one propagator, as we did for the scalar
coupling.

The gauge coupling of the effective 4D theory is at the one loop level
given by
\begin{equation}\label{gL}
g^{-2}=g_0^{-2}[1+g_0^2 \beta_0 \log(\mu_{ir}R)-g_0^2 \beta_1\log(\Lambda R)],
\end{equation}
\noindent where
\begin{equation}
\frac{\beta_0}{2}=\beta_1=\frac{1}{48\pi^2}.
\end{equation}
\noindent In eq.~(\ref{gL}), as in eq.~(\ref{LoopSZ}) for the one-loop
scalar coupling, the $\log(\Lambda R)$ comes from brane effects.

If the gauge group is non-abelian, then we only have to modify the
charges and multiply by $t(S)$, where $tr[T_a(S)T_b(S)]=t(S)\de_{ab}$.

Eq.~(\ref{gL}) gives the scalar contribution to the 4D effective
coupling at one loop, for a theory with a
cut-off scale $\Lambda$. We can consider a theory with a different
cut-off $\Lambda'$, and one-loop coupling $g'$. Then the relation
between the couplings $g$ and $g'$ is given by
\begin{equation}
g'=g[1+g_0^2 \beta_1\log\frac{\Lambda'}{\Lambda}].
\end{equation}

\section{6D winding renormalization}
\begin{figure}[htbp]
    \centering 
    \psfrag{R1}{$2\pi R_1$}
    \psfrag{2R1}{$4\pi R_1$}
    \psfrag{R2}{$2\pi R_2$}
    \psfrag{2R2}{$4\pi R_2$}
    \psfrag{y1}{$y_1$}
    \psfrag{y2}{$y_2$}
    \psfrag{y}{$\bar{y}$}
    \psfrag{t}{$\theta$}
    \includegraphics[width=7cm]{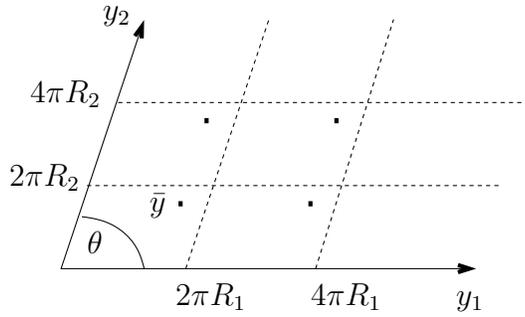}    
    %\end{picture}
    \caption{\textit{Identification of $\mathcal{R}^2$ to obtain a
      bidimensional torus $T^2$, with parameters $\{R_1,R_2,\theta\}$.}}
    \label{id6D}
\end{figure}

We apply winding modes formalism to a space with two
extra-dimensions. Given an infinite plane $\mathcal{R}^2$ we can obtain
a two-dimensional torus $T^2$ identifying with
$G=\mathcal{Z}\times\mathcal{Z}$ ($\mathcal{Z}$ the integer numbers). The identification is
$\bar{y}\sim\bar{y}+\bar{w}$, where $\bar{y}=(y^1,y^2)$ and
$\bar{w}=2\pi(n^1R_1,n^2R_2)$, $n^j\in\mathcal{Z}$. $R_1$ and $R_2$
meassure the size of the extra dimensions compactified in a
torus, $y_j\in[0,2\pi R_j)$. There is one parameter more to obtain a
complete description of $T^2$, the angle betwen the directions of
identification in the plane, as shown in fig.~\ref{id6D}.

In the same way as in 5D we can find the scalar Green's function in euclidean
infinite space, in mixed representation (momenta in Minkowski
directions and coordinates in extra directions)
\begin{equation}
(p^2-\partial^j \partial_j)\G(p;\bar{y}-\bar{y}')=\delta^{(2)}(\bar{y}-\bar{y}'),
\end{equation}
\noindent where $p^2=p^{\mu}p_{\mu}$ is the 4D momentum and $j=1,2$
numbers the extra dimensions. The solution to this equation is
\begin{equation}
\G(p;\bar{y}-\bar{y}')=\frac{1}{2\pi}K_0(p|\bar{y}-\bar{y}'|),
\end{equation}
\noindent where $K_0$ is $K$ Bessel function of zero order. Now we
identify and get the propagator in the compact space
\begin{equation}
\G^{tor}(p;\bar{y},\bar{y}')= \sum_{\bar{w}} \frac{K_0(p|\bar{y}-\bar{y}'+\bar{w}|)}{2\pi},
\end{equation}
\noindent with $|\bar{y}|$ the modulus of the vector messured with the
flat metric of a torus
\begin{equation}
ds^2=dy_1^2+dy_2^2+2\cos^2 \theta dy_1 dy_2.
\end{equation}

As in the 5D case, the propagator for non-zero winding is
exponentially damped at high energies $p|\bar{w}|\gg 1$
\begin{equation}
K_0(p|\bar{w}|)\rightarrow e^{-p|\bar{w}|}\sqrt{\frac{\pi}{2p|\bar{w}|}},
\end{equation}
\noindent showing that winding contributions will be always finite.

For arbitrarily small argument $K_0(x)\rightarrow -\log x$. Then the
propagator diverges at short distances in the extra
dimensions. Therefore to compute Feynman integrals with
$\bar{y}\rightarrow 0$ we have to regulate the propagator\footnote{In D-dimensions with two of them in coordinate
  representation and D-2 in momentum representation, the propagator in flat
  infinite space is $\G^D(p;\bar{y})=\int
  d^2s\frac{e^{i\bar{s}.\bar{y}}}{p^2+s^2}$, where we see that
  evaluating $\bar{y}=0$ we get logarithmic divergences. We can
  regulate the short distance behaviour with an UV cut-off
  $|\bar{y}|_{min}=\Lambda^{-1}$, then the propagator becomes
  $\G^{\Lambda}(p;0)=\pi\log(\frac{p^2+\Lambda^2}{p^2})$.} when
$\bar{w}=0$.

To compactify on an orbifold we have to introduce new
identifications. A simple possibility is obtained introducing two $Z_2$
groups, one acting on $y^1$ and the other on $y^2$, in this way
$y^j\sim -y^j$. Then, due two $\mathcal{Z}$ and $Z_2$ action on each
direction, the fundamental domain becomes $[0,\pi R_j]$. We can act
with each $Z_2$ independently, then we identify four different points
on $\mathcal{R}^2$: $(y^1,y^2)\sim (\pm y^1,\pm y^2)$. According to
this, the orbifold 6D propagator is
\begin{equation}\label{GTZ2}
\begin{aligned}
\G^{orb}(p;\bar{y},\bar{y}')= \sum_{\bar{w}}&[\G(p;\bar{y}-\bar{y}'+\bar{w})+p_1\G(p;\bar{y}-\bar{x}'+\bar{w})\\&+p_2\G(p;\bar{y}+\bar{x}'+\bar{w})+p_1p_2\G(p;\bar{y}+\bar{y}'+\bar{w})],
\end{aligned}
\end{equation}
\noindent where $\bar{x}'=(-y'^1,y'^2)$ and $p_j$ is the field parity in
$j$ direction. 

Here we can make the
same analysis as in 5D: the new propagator terms will give localized
divergent contributions. For zero winding the terms with just one
coordinate identified under $Z_2$ will give 5D divergencies (localized in
one extra dimension), and the term with both coordinates inverted will give
4D divergencies localized in a point of the extra space.

\subsection{6D scalar renormalization on $T^2$}
Using the winding modes we can easily separate cut-off dependent from finite
contributions in 6D. To show this let's compute the radiative corrections
of the scalar theory at one-loop with the extra space compactified on
a torus $T^2$. The effective action is similar to eq.~(\ref{Seff}) but
with two extra dimensions. For a torus $m^2(y)$ is constant, and is
given by
\begin{equation}\label{masaT}
m^2_{tor}=\frac{\lambda}{2}\int \intp \G(p,\bar{w})=\frac{\lambda}{32\pi^3}(\frac{\pi^2 \Lambda^4}{2}+\sum_{\bar{w}\neq0} \frac{4}{|\bar{w}|^4}).
\end{equation}
Again the divergent term is due to the zero winding
mode and winding modes different from zero give the finite contributions. 

Let's consider a constant field $\phi_c$. Integrating over the extra space we
can compare the finite 4D masses obtained from a 5D theory with the
ones obtained from a 6D
theory. Making $\lambda_{5D}=\lambda_{6D}/2\pi R$ and summing over
windings we get $m^{cir}_{4D}\simeq 1.5\;m^{tor}_{4D}$.

\begin{figure}[htbp]
    \centering 
    \psfrag{1R1}{$2\pi R_1$}
    \psfrag{1R2}{$2\pi R_2$}
    \psfrag{y1}{$w_1$}
    \psfrag{y2}{$w_2$}
    \includegraphics[width=4cm]{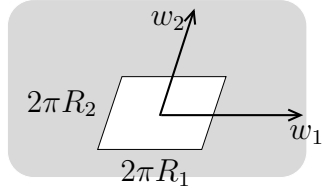}    
    %\end{picture}
    \caption{\textit{We consider $\bar{w}$ a continuos variable, then
    it takes values over the plane. As we have separated the zero mode,
    we don't integrate over the parallelogram in the origin.}}
    \label{guraco}
\end{figure}
Now we consider $S_4$ for a constant field $\phi_c$. Using the analog
to equation (\ref{simple2}) but with one more extra dimension we can
write the 6D one-loop coupling as 
\begin{equation}\label{verticecorto}
\lambda_{tor}=-\frac{\lambda^2}{2} \sum_{\bar{w}}\int \intp
\,\frac{d}{dp^2} \G(p;\bar{w})=
\frac{\lambda^2}{32\pi^3}(\frac{\Lambda^2}{4}+\sum_{\bar{w\neq 0}}\frac{1}{|\bar{w}|^2}),\;\;\;\;\;\;\;\;\;\;\phi=\phi_c.
\end{equation}
The sum over modes is
logarithmically divergent, in the same way as in 5D. We approximate
the sum with an integral, then $\bar{w}\in R^2$. We have to exclude
the zero mode, then the domain of intagration is shown in
fig.~\ref{guraco}. 
Therefore the second term of the r.h.s. of eq.~(\ref{verticecorto}) is
\begin{equation}
\simeq
-\frac{\lambda^2}{16\pi^3V_{tor}}[(\pi-\gamma)\log(\mu_{ir} R_1)+\gamma\log(\mu_{ir} R_2)],\;\;\gamma=arctg(\frac{2R_1 R_2 \sin\theta}{R_1^2-R_2^2}),
\end{equation}
\noindent where $\mu_{ir}$ is an IR cut-off, the inverse of
$|\bar{w}_{max}|$ and $V_{tor}$ is the torus volume.

\subsection{6D scalar renormalization on an orbifold $T^2/Z_2\times Z_2$}
We repeat the steps done for the 6D torus, using the orbifold 6D
propagator of eq.~(\ref{GTZ2}). For simplicity we consider a constant
field $\phi_c$. Then we can write $S_2$ as
\begin{equation}\label{resultadomT/Z2}
S_2=-\phi_c^2\big\{m^2_{tor}V_{orb}+m^2_f+\frac{\lambda}{32\pi^3}\big[\sum_{j=1,2}\frac{p_jV_{orb}\Lambda^3}{6\pi
  R_j} +\frac{p_1p_2}{2}\Lambda^2\theta\big]\big\},\;\;\;\;\;\;\;\;\;\;\phi=\phi_c
\end{equation}
\noindent where $V_{orb}$ is the orbifold volume and $m^2_f$ is a
  finite term given by
\begin{equation}
\begin{aligned}\label{mf}
m^2_f=&\frac{\lambda}{32\pi^3}\big\{\sum_{j=1,2}\frac{-p_j V_{orb}}{6\pi^4
  R_j^4}-\frac{p_1p_2}{8\pi^2\sin^2\theta}[\frac{f(\beta,\theta-\beta)-\sin 2\theta}{R_1^2}-\frac{f(\theta-\beta,\beta)-\sin 2\theta}{R_2^2}]\big\}
\\
&+\frac{\lambda}{32\pi^3}\int \!\!
  d^2y\sum_{\bar{w}_{fin}}[\frac{p_1}{|2(y^1,0)+\bar{w}|^4}+\frac{p_2}{|2(0,y^2)+\bar{w}|^4}+\frac{p_1p_2}{|2\bar{y}+\bar{w}|^4}],
\end{aligned}
\end{equation}
\noindent the function $f(x,y)$ is defined by
\begin{equation}
f(x,y)=2x+\sin 2y,\;\;\;\;\;\;\;\;\sin^2(\beta)=\frac{(R_2 \sin\theta)^2}{(R_1)^2+(R_2)^2+2 R_1 R_2 \cos\theta}.
\end{equation}
\noindent The sum in eq.~(\ref{mf}) is over windings that
  do not give divergencies. Then we have to exclude windings given by
$(n^1,n^2)=(0,0),(-1,0),(0,-1),(-1,-1)$, when these windings give
divergencies.

The $\Lambda^3$ divergencies in eq.~(\ref{resultadomT/Z2}) are
localized in one direction, this can be seen considering fields
$\phi(\bar{y})$, and expanding them in
power series around the fixed points. This
divergencies are similar to the bulk divergencies in 5D. Using the
series expansion it can be seen that
$\Lambda^2$ terms are 4D, they are localized on the four
fixed points $\bar{y}_{fp}$.
To obtain divergent localized kinetic terms in the
directions of the extra dimensions, we have to consider the terms of
second order in the series expansion.

Now we consider $S_4$ for the orbifold with constant
$\phi_c$. Integrating (\ref{simple2}) in 6D with the
orbifold propagator we obtain $S_4$ at one-loop. We approximate the
sum over windings with an integral. The result depends on the $Z_2$ parity
$p_j$, and is given by
\begin{equation}
\begin{aligned}
S_4\simeq\frac{\lambda^2\phi_c^4}{32\pi^3}&\{\frac{\Lambda^2
  V_{orb}}{4}+\sum_{j=1,2}\frac{\Lambda p_j \pi R_j\sin\theta}{8}+2
p_1p_2[\beta\log(\Lambda R_1)+(\theta-\beta)\log(\Lambda R_2)]
\\
&-(1+p_1+p_2+p_1p_2)[\frac{\pi-\gamma}{2}\log(\mu_{ir} R_1)+\frac{\gamma}{2}\log(\mu_{ir} R_2)]\},
\end{aligned}
\end{equation}
\noindent where we have regularized the winding sum with an IR cut-off $\mu_{ir}$. 

The IR logarithmic contributions are cancelled if the scalar field is
odd in any of the directions. This again is easier to understand with
K-K decomposition, there is zero mode just for the even-even case.
Evaluating this equation for $\{R_1=R_2=R, \theta=\pi/2\}$ is very
easy to compare the linear and logarithmic divergencies with the 5D
case.

The radiative corrections show that we should include bulk, 5D
and 4D localized masses from the begining, and also
localized vertices and kinetic terms.

\section{Radion stabilization in plane orbifolds}
%The hierarchy problem can be solved with $n$ large extra dimensions
%\cite{add}. The key observation is that gravitation is weak because
%it propagates in the extra dimensions, while the SM fields are
%localized on a 3-brane. As a consequence of this, if the space is
%approximately a direct product, the Planck scale $M_{Pl}$ is related
%to the fundamental mass scale $M$ (of order $m_{EW}$ the electroweak
%scale) of the higher dimensional theory by $M_{Pl}^2= M^{2+n} V_n$,
%where $V_n$ is the volume of the extra space. To build a realistic
%theory in this scenario it is necesary that $n\geq 2$ and
%$MV_n^{1/n}\gg 1$, thus we need to stabilize a big volume.  One
%posibility for stabilization is the Cassimir energy
%\cite{WeinbergCandelas}, there have been many attempts in this
%direction but almost all of them demand fine tunnings. In reference
%\cite{alexcasimir} stabilization is obtained with gauge fields in RS scenario.

As an application of the winding
formalism, we compute in this section the leading two loop contributions to the
effective potential for the radion, in a product space
$\mathcal{R}^4\times S^1/Z_2$. We will see that under certain
symmetry assumptions we can get a Coleman-Weinberg
potential. Therefore the size of the extra dimension can be stabilized
at large values.

%Although our scenario has just one extra dimension, it shows the way of thinking to proceed with more realistic theories.

\subsection{Scalar potential}
\begin{figure}[htbp]
    \centering 
    \psfrag{a}{(a)}
    \psfrag{b}{(b)}
    \psfrag{c}{(c)}
    \includegraphics[width=7cm]{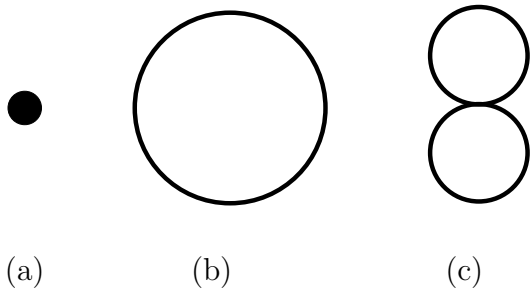}
    \caption{\textit{Perturbative expansion of the effective potential
    for the radion. The first diagram (a) is the tree level
    contribution, that cancels out.}}
\label{Figloopsescalares}
\end{figure}

Let's consider a scalar 5D theory, as the one in section~\ref{orbren},
compactified on an orbifold. The effective potential for the
radion at tree
level is zero, so we compute loops to obtain a sensible effective
potential (see fig.~\ref{Figloopsescalares})~\footnote{If the scalar
  vev $\langle\phi\rangle\neq 0$, we also have to include a two loop
  diagram, with two three-point vertices, each of them proportional to the
  vev~\cite{V2loops}.}. We calculate these
quantum corrections using the winding formalism. Let's start with the
one-loop term.
In K-K modes we can write the one loop effective potential as
\begin{equation}
V^{(1)}=\sum_{k_5} \int \dk \log(k^2+k_5^2).
\end{equation}
\noindent To obtain it in winding representation we can write the last
equation as
\begin{equation}
V^{(1)}=\sum_{k_5} \int \dk \int dk^2\frac{1}{(k^2+k_5^2)}. 
\end{equation}
\noindent The last factor is the scalar propagator in K-K modes, then
we can replace it by the one with winding modes and integrate
\begin{equation}
\begin{aligned}
V^{(1)}&=\sum_n \int \dk \int dk^2 \int_0^{\pi R}\!dy\; [\G(k,2n\pi
R)\pm\G(k,2y+2n\pi R)]
\\ &=\frac{1}{8\pi^2}\big[\frac{\Lambda^5}{5}\pi R-\frac{3\zeta(5)}{8\pi^4 R^4}\mp\frac{\Lambda^4}{16}\big],
\end{aligned}
\end{equation}
\noindent where $\G$ is defined in eq.~(\ref{GR1}), $\zeta$ is the
Riemmann zeta function
($\zeta(\alpha)=\sum_{n>0} 1/n^{\alpha}$), and we get the finite term
from no zero windings.

After that we compute the two loop term $V^{(2)}$, shown in (c) of
fig.~\ref{Figloopsescalares}. It is given by
\begin{equation}\label{V2top}
  \begin{aligned}
    V^{(2)}=\frac{\lambda}{2} \sum_{n,n'} \int_0^{\pi R}\!\!\!\! dy \int \dk \dq [ & \G(k,2n\pi
    R)\pm\G(k,2y+2n\pi R)]\\ \times & [\G(q,2n'\pi R)\pm\G(q,2y+2n'\pi R)].
\end{aligned}
\end{equation}
Once more zero winding modes are divergent and non zero modes give
the finite contributions. If we call $\G_S$ the first
term of the orbifold propagator (identic to the circle
propagator) and $\G_Z$ the second term (obtained with $Z_2$
identification), then we can write equation (\ref{V2top}) as
$V^{(2)}=V_{SS}\pm 2V_{SZ}+V_{ZZ}$. Let's consider first the term
$V_{SS}$. Every loop is similar to the one loop contribution to the
mass.
% then if we just integrate coordinates and momenta, we can write
%$V_{SS}$ as 
%\begin{equation}\label{VSS}
%V_{SS}=\frac{\lambda \pi
%R}{(8\pi^2)^2}[\frac{\Lambda^3}{6}+\frac{\zeta(3)}{8\pi^3
%  R^3}]^2.
%\end{equation}
%\noindent %there is a factor $\pi R$ corresponding to the orbifold volume.
%But this conclusion is wrong, due to the cross term, proportional to
%$\Lambda^3 R^{-2}$. To understand why we have to realize the following
%thing: the term 
Then $V_{SS}$ of eq.~(\ref{V2top}), can be interpreted as one
loop with a massless propagator $\G_{S}$ and a mass
$m^2_{cir}\propto\lambda[\Lambda^3/6+\zeta(3)/(8\pi^3 R^3)]$, as is
shown in the following Feynman diagram
\begin{figure}[htbp]
    \centering
    \psfrag{V}{$V_{SS}$}
    \psfrag{=}{$=$}
    \psfrag{+}{$+$}
    \psfrag{m}{$m^2_{cir}$}
    \includegraphics[width=6cm]{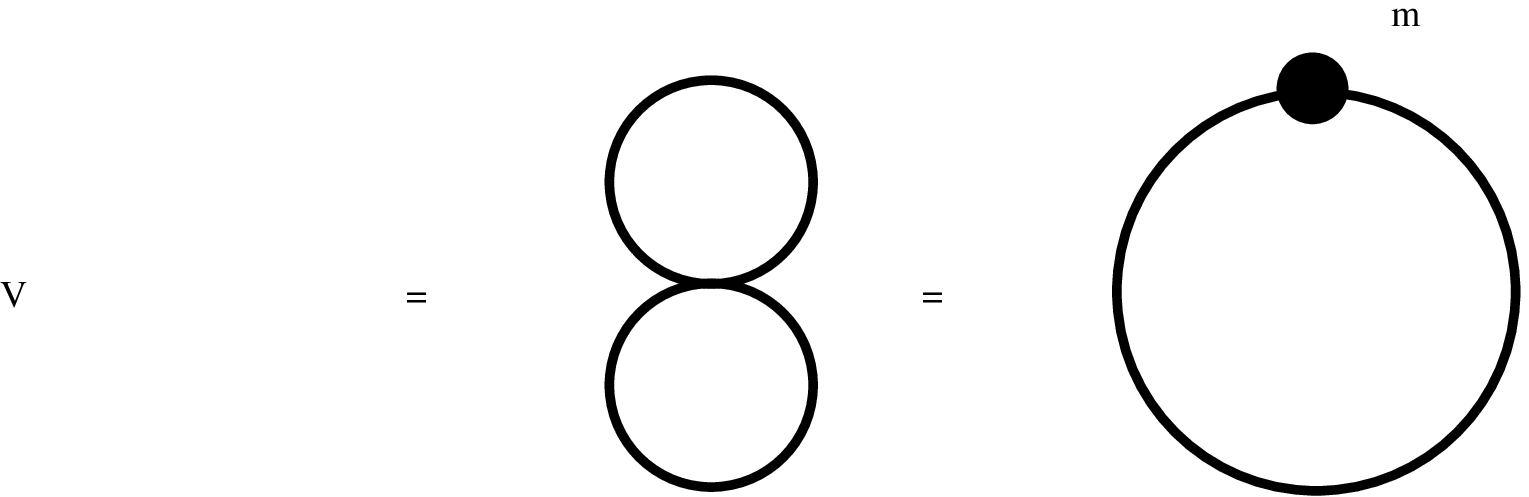}
    %\caption{\textit{}}
\label{VSSmasa}
\end{figure}

\begin{figure}[htbp]
    \centering
    \psfrag{+}{$+$}
    \psfrag{l}{$m^2_{cir}$}
    \psfrag{G}{$\G_{massive}$}
    \psfrag{+...=}{$+\;\;\;...\;\;=$}
    \includegraphics[height=2.6cm]{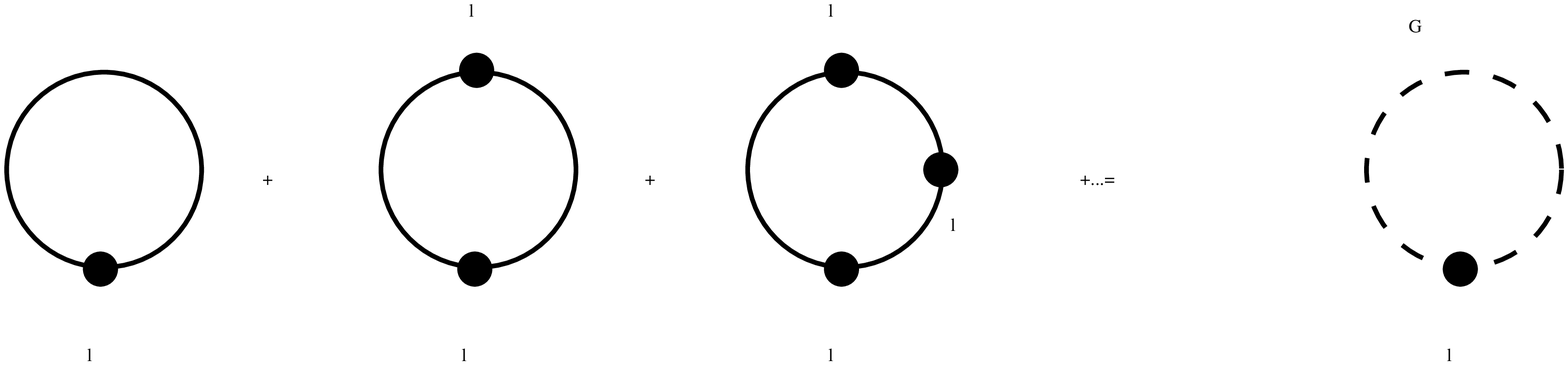}
    \caption{\textit{Feynman diagrams giving the one loop effective
    potential with mass insertions. The mass $m_{cir}$ itself is one
    loop, it is
    given by the one loop contribution to the mass. The continuos lines
    are for masless propagators and the dashed line is for the massive one.}}
\label{FigGs2}
\end{figure}

\noindent The mass $m_{cir}$ is of order $\Lambda$, then it can not be considered a
perturbation. Therefore we have to
consider terms with arbitrary number of mass insertions, as is shown
in fig.~\ref{FigGs2}.
If we sum the series of fig.~\ref{FigGs2}, we
obtain a loop with a massive propagator and a mass vertex.
As we discussed in section~\ref{windings}, a massive propagator
is obtained replacing $p\rightarrow \sqrt{p^2+m^2_{cir}}$. If the mass
is large, $m_{cir}\gg R^{-1}$ (as is the case because $m_{cir}\sim\Lambda\gg R^{-1}$), we can approximate the propagator by $e^{-m_{cir}|y+2n\pi
R|}/m_{cir}$. In this case the propagator is exponentially damped and
cancels before making windings. Then the only relevant contributions
are divergent, and $V_{SS}\simeq 2\lambda \Lambda^6\pi R/(96\pi^2)^2$.

If there is a symmetry prohibiting divergent masses, as a local
gauge symmetry, then the finite contributions are given by
$V_{SS}\simeq 2\lambda \zeta(3)^2/(128^2\pi^9 R^5)$.

We can make a similar analysis for the other topologies, obtaining
the same results. Summing over topologies we get
\begin{equation}\label{V2f}
V^{(2)}=\frac{\lambda}{64 \pi^4}[A\; (\pi R)\Lambda^6+B
\Lambda^5+C(\pi R)^{-5}],
\end{equation}
\noindent where $A\sim 10^{-1}, B\sim 10^{-1}, C\sim 10^{-2}$, and
the sign of $B$ depends on field parity under $Z_2$. If there is
no symmetry protecting $V$ from divergencies, the divergent terms are
dominant, in the other case we only get the $R^{-5}$ finite term.

\subsection{Effect of brane kinetic terms}
\begin{figure}[htbp]
    \centering
    \psfrag{0}{$0$}
    \psfrag{p}{$\pi R$}
    \psfrag{ly}{$\lambda_y$}
    \includegraphics[width=5cm,height=3cm]{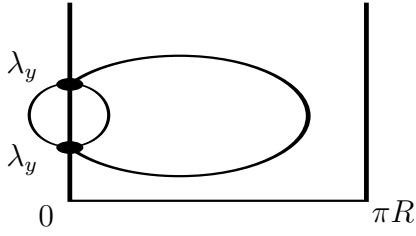}
    \caption{\textit{Feynman diagram with one localized fermionic loop and one
    bulk scalar loop, contributing to the effective potential.}}
\label{figloopferm}
\end{figure}

We want to obtain a potential able to stabilize a large extra dimension. 
Then we consider new brane terms: we add to the previous set-up fermion
fields localized on the fixed points.
If these 4D fields couple to the bulk ones, there are new
contributions to the two loop effective potential. There is a new term
with a fermionic-loop localized on the branes and a scalar-loop on the
bulk, as is shown in fig.~\ref{figloopferm}.
To be more precise we consider the following interaction
$\mathcal{L}_y=\de(y)\lambda_y \phi \bar{\psi} \psi$, localized on
the brane at $y_{fp}=0$, and compute the two loop contribution.

The 4D fermionic loop has several terms given by
\begin{equation}\label{loop4D}
\frac{\lambda_y^2}{8\pi^2}
[-2\Lambda^2+\frac{5}{3}q^2-q^2\log \frac{q^2}{\Lambda^2}].
\end{equation}
\noindent We are
interested in the logarithmic kinetic contributions. The first
and second terms are local, but the third is not local and will give
us new things.

The term of the scalar effective action that couples to the brane
loop is
\begin{equation}
\frac{\lambda_y^2}{8\pi^2}\int_0^{\pi R}\!\!\!\! dy\;\int_0^{\pi
R}\!\!\!\! dy'\; \int \intp \de(y)\de(y') \phi(y) p^2
\log(\frac{p^2}{\Lambda^2}) \phi(y').
\end{equation}
\noindent Then, the two loop contribution with a loop localized on one
of the branes, is given by
\begin{equation}\label{V2w}
\begin{aligned}
V^{(2)}_b &=\frac{\lambda_y^2}{8\pi^2} \sum_n\int \intp
\G(p;2n\pi R) p^2\log(\frac{p^2}{\Lambda^2})
\\ &=\frac{\lambda_y^2}{64\pi^4}[\frac{\Lambda^5}{25}+\sum_{n\neq
  0}\frac{50-24 \gamma-24\log(2n\pi R\Lambda)}{(2n\pi R)^5}],
\end{aligned}
\end{equation}
\noindent as usual, the divergent term is due to the zero winding.

Now we have to sum $V^{(1)}+V^{(2)}+V^{(2)}_b$ to get the effective
potential to two loops. Let's suppose that for each boson there is a
fermion with equal boundary conditions. Then the finite
part of $V^{(1)}$
cancels out, because fermionic loops have a minus sign. Furthermore,
if there is a local symmetry (like local supersymmetry) protecting the
effective potential from divergent terms, then only the finite terms in
$V^{(2)}$ and $V^{(2)}_b$ remain, and the two-loop effective potential
is given by
\begin{equation}\label{Rgrande}
V\sim \frac{1}{R^5}[cte.-\log(\Lambda R)], \;\;\;\;\;\;\;\; cte. \sim 1.
\end{equation}
\noindent This is a Coleman-Weinberg potential \cite{cw}. We know that
this kind of potential can stabilize $R$ with a large value $R\gg
\Lambda^{-1}$, that is consistent with our renormalization
assumptions and the large extra dimensions scenario. In this way
quantum corrections in a product space with matter fields
localized on the branes can stabilize the size of the extra space,
with large radius. 

\section{Conclusions}
We have used the winding formalism to compute radiative corrections on
a theory with extra dimensions. It allows us to
separate, in a very clear and intuitive way, cut-off dependent from finite
corrections. It is also very easy to see how the brane terms are
generated at one loop. We extended this formalism to a
higher dimensional space, and showed that it is immediate to
separate finite from cut-off dependent contributions, making this
method very useful.

We explored the effects of parallel and perpendicular kinetic brane
terms. Our conclusions are that, whenever there are perpendicular
kinetic terms on the branes, the theory can be redefined, in such a
way that there only remain parallel kinetic terms.
%(at low energies: $E\ll \Lambda$, where the theory is well defined). 
We also argued that brane kinetic terms can provide a new way of
symmetry breaking.

We applied the winding formalism to compute the finite terms of scalar
masses, in theories with one
and two extra dimensions. If there is a symmetry (like supersymmetry, or other
global or gauge symmetries) that protects masses from
divergencies, this finite terms are predictions of the theory. We also
get, in theories with one and two extra dimensions, the logarithmic
contributions to the 4D couplings in a simple way.

We analyzed the possibility of getting a potential
stabilizing the size of the extra space, when the higher dimensional
space can be approximated by a direct product. We saw that it is
very simple to compute the two loop effective potential with winding
modes. We also showed an scenario with brane terms and bulk fields
where the extra volume can be stabilized with a Coleman-Weinberg
potential.

\section{Aknowledgments}
I would like to thank {\'A}lex Pomarol for many ideas, discussions and
constant advise. I also acknowledge fruitful discussions with
A. Flachi and O. Pujol{\`a}s.
This work was supported by the Spanish Education Office (MECD) under
an FPU scolarship.
%====================================================================

\end{document}